\documentclass[prd,preprint,floatfix,nofootinbib,superscriptaddress,showpacs]{revtex4} 

\pdfoutput=1
\usepackage{epsfig}
\usepackage[caption=false]{subfig}
\usepackage{amsmath}
\usepackage{amsfonts}
\usepackage{amssymb}
\usepackage{float}
\usepackage{color}
\usepackage{graphicx}
\usepackage{graphics}
\usepackage{hyperref}
\usepackage{adjustbox,array}
\hypersetup{}
\usepackage[utf8x]{inputenc} 
\usepackage{epstopdf}
\usepackage{multirow}
\usepackage{slashed}
\usepackage{booktabs}
\definecolor{rosso}{cmyk}{0,1,1,0.4}
\definecolor{rossos}{cmyk}{0,1,1,0.55}
\definecolor{rossoc}{cmyk}{0,1,1,0.2}
\definecolor{blu}{cmyk}{1,1,0,0.3}
\definecolor{blus}{cmyk}{1,1,0,0.6}
\definecolor{bluc}{cmyk}{1,1,0,0.1}
\definecolor{verde}{cmyk}{0.92,0,0.59,0.25}
\definecolor{verdec}{cmyk}{0.92,0,0.59,0.15}
\definecolor{verdes}{cmyk}{0.92,0,0.59,0.4}

\AtBeginDocument{
\heavyrulewidth=.08em
\lightrulewidth=.05em
\cmidrulewidth=.03em
\belowrulesep=.65ex
\belowbottomsep=0pt
\aboverulesep=.4ex
\abovetopsep=0pt
\cmidrulesep=\doublerulesep
\cmidrulekern=.5em
\defaultaddspace=.5em
}

\hypersetup{
 colorlinks=true,
 linkcolor=magenta,
 urlcolor=magenta,
 citecolor=magenta
}

\newcommand{\orcidAJ}{\href{https://orcid.org/0000-0002-0963-366X}{\includegraphics[height=9pt]{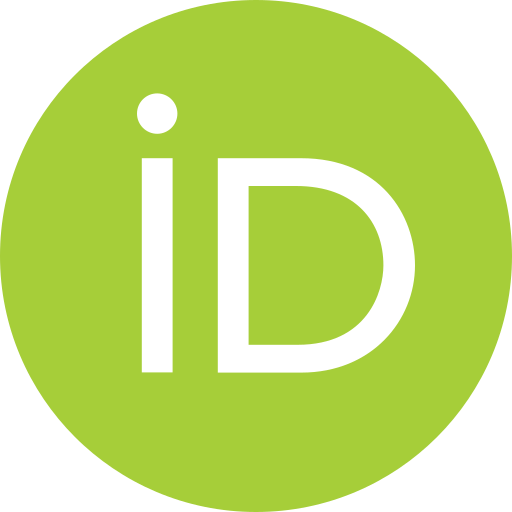}~}}
\newcommand{\orcidSN}{\href{https://orcid.org/0000-0002-5985-4567}{\includegraphics[height=9pt]{Figures/Logos/orcid.png}~}}

\begin{document}

\title{\color{blue} Phenomenology of Minimal Leptophilic Dark Matter Models at Linear Colliders}

\author{Adil Jueid\footnote{Speaker}\footnote{Talk presented at the International Workshop on Future Linear Colliders (LCWS 2021), 15-18 March 2021, PD1.}\orcidAJ}
\email{adil.hep@gmail.com}
\address{Department of Physics, Konkuk University, Seoul 05029, Republic of Korea}

\author{Salah Nasri\orcidSN}
\email{salah.nasri@gmail.com}
\address{Department of physics, United Arab Emirates University, Al-Ain, UAE}
\address{The Abdus Salam International Centre for Theoretical Physics, Strada
Costiera 11, I-34014, Trieste, Italy}

\begin{abstract}
We discuss the phenomenology of a minimal model for GeV-scale Majorana dark matter (DM) coupled to the standard model lepton sector via a charged scalar singlet. The theoretical framework extends the Standard Model by two $SU(2)_L$ singlets: one charged Higgs boson and a singlet right-handed fermion. The latter plays the role of the DM candidate. We show that there is an anti-correlation between the spin-independent DM-Nucleus scattering cross-section ($\sigma_{\rm SI}$) and the DM relic density. We suggest a scenario that can be tested in both electron-electron and electron-positron collisions using the production of same-sign charged Higgs pairs and mono-Higgs in association with right-handed neutrinos.
\end{abstract}


\maketitle

\section{Introduction}
\label{sec:intro}
The search for particle dark matter (DM) in the universe is still ongoing in direct detection \cite{Aprile:2018dbl, Cui:2017nnn}, indirect detection \cite{Adriani:2010rc, Aguilar:2013qda, Ahnen:2016qkx, Abdallah:2016ygi} and collider experiments \cite{Meirose:2016pxn, Ahuja:2018bbj}. Despite these tremendous efforts, only few mild excesses have been reported whose statistical significances did not cross the $5\sigma$ threshold\footnote{In this kind of experiments, even a statistical significance above $5\sigma$ is not sufficient to claim a discovery due to the possibility of {\it e.g.,} incomplete understanding of the backgrounds, errors in the calibrations...etc.}. These null results may be considered as disappointing for Weakly Interacting Massive Particles (WIMPs) as it implies: {\it (i)} the DM particle(s) couple super-weakly to the Standard Model (SM) sector or {\it (ii)} relatively light DM candidates with masses $M < \mathcal{O}(10^3)~{\rm GeV}$ are disfavored. From the absence of strong signals in indirect detection experiments, most of the minimal dark-matter models where the annihilation cross section occurs through $s$-wave amplitudes are excluded. Another problem is that the production cross section of DM at colliders is usually strongly correlated to the DM-nucleus scattering cross sections for minimal dark-matter models\footnote{To avoid this, minimal models should be extended by various new states to allow for some cancellations in the predicted cross sections. We notice that the drawback of this non-minimal extension is the loss of the predictive power.}. This has led to the exclusion of the most minimal model for GeV-scale DM candidate, {\it i.e.,} the SM with real scalar singlet \cite{Silveira:1985rk}. The situation is different if the dark-matter is a singlet Majorana fermion. The reasons for this are: {\it (i)} the annihilation cross section is dominated by $p$-wave amplitudes which is velocity suppressed, {\it (ii)} the scattering of the singlet Majorana fermion off the nucleus is always induced at the one-loop order due to the absence of tree level couplings of DM to $Z/H$ bosons and {\it (iii)} constraints from LHC searches of dark-matter are usually weak if the dark matter couples only to charged leptons. Models based on singlet Majorana fermions are phenomenologically appealing as they can be extended to accommodate for the smallness of neutrino masses via radiative neutrino mass generation (see {\it e.g.,} \cite{Ma:1998dn, Krauss:2002px}). In this talk, we discuss the phenomenology of a very minimal model which extends the SM particle content with two gauge singlets; a charged scalar and neutral Majorana fermion \cite{Jueid:2020yfj}. Probes of this model through gravitational waves have been recently studied by the authors of \cite{Liu:2021mhn}. The remainder of this article is organized as follows. In section \ref{sec:model}, we discuss the theoretical setup of the model. We discuss all the theoretical and the experimental constraints on the model parameter space in section \ref{sec:constraints}. In section \ref{sec:DM}, we discuss the dark-matter phenomenology of the model. The implications of the model at the International Linear Collider are discussed in section \ref{sec:ILC}. We conclude in section \ref{sec:conclusion}.

\section{The model}
\label{sec:model}
We discuss briefly in this section the theoretical setup of our model. Here, we consider a minimal extension of the SM by 
\begin{eqnarray}
{\rm Charged~scalar~singlet}~(S): \quad ({\bf 1, 1, 2}, -1), \nonumber \\
{\rm Singlet~Majorana~fermion}~(N_R): \quad ({\bf 1, 1, 0}, -1), 
\label{eq:rep}
\end{eqnarray}
where the numbers inside the parentheses show the transformations of these new states under $SU(3)_c \otimes SU(2)_L \otimes U(1)_Y \otimes Z_2$. It is clear from eqn. (\ref{eq:rep}) that $S$ and $N_R$ are odd under $Z_2$ so that the lightest neutral odd particle (LNOP), here $N_R$, would be the DM candidate in our model. We must note that one could assign a lepton number to the charged scalar singlet $S$\footnote{The charged scalar singlet is also called scalar lepton \cite{Baum:2020gjj}.}. This implies that DM is considered as leptophilic as it only couples to the charged SM leptons. Therefore, the most general renormalizable Lagrangian is given by
\begin{eqnarray}
\mathcal{L} = \mathcal{L}_{\rm SM} - \mathcal{L}_{N_R} + \mathcal{L}_{\rm gauge} - V(\Phi, S).
\label{eq:Lag:tot}
\end{eqnarray}

In Eqn. (\ref{eq:Lag:tot}), $\mathcal{L}_{N_R}$ represent the Lagrangian involving the charged lepton, the singlet fermion, and the singlet charged scalar:
\begin{eqnarray}
	\mathcal{L}_{N_R} \supset \sum_{\ell} y_\ell \bar{\ell}^c_R S N_R + \frac{1}{2} M_N \bar{N}_R (N_R)^c + h.c.
\label{eq:int:1}	
\end{eqnarray}

The most general renormalizable, gauge-invariant and {\it CP}-conserving scalar potential is given by 
\begin{eqnarray}
V(\Phi, S) &=& - m_{11}^2 |\Phi^\dagger \Phi| + m_{22}^2 |S^\dagger S| + \lambda_1 |\Phi^\dagger \Phi|^2  + \lambda_2 |S^\dagger S|^2 + \lambda_3 |\Phi^\dagger \Phi| |S^\dagger S|,
\end{eqnarray}
The SM Higgs isodoublet can be parameterised, in the unitary gauge, as $\Phi^T = \frac{1}{\sqrt{2}}(0, v+H)$. The $CP$-invariance of the model implies that $m_{11}^2, m_{22}^2, \lambda_{i,i=1,2,3}$ are real parameters.\\

The gauge interaction of the charged scalar singlet is given by
\begin{eqnarray}
\mathcal{L}_{\rm gauge} = (\mathcal{D}^\mu S)^\dagger (\mathcal{D}_\mu S) \supset - i(e A^\mu - e \tan\theta_W Z^\mu) S^\dagger \bar{\partial}_\mu S, 
\label{eq:int:2}
\end{eqnarray}
with $A\bar{\partial}B = (\partial A)B - A(\partial B)$. After electroweak symmetry breaking, we are left with one CP-even neutral scalar $H$ which we identify with the SM Higgs boson and a pair of charged scalars ($H^\pm$): 
$$
m_H^2 = \lambda_1 v^2, \quad m_{H^\pm}^2 = m_{22}^2 + \frac{1}{2} \lambda_3 \upsilon^2
$$
The model involves additionally seven new parameters: two masses ($m_N$ and $m_{H^\pm}$), three Yukawa couplings ($y_e, y_\mu$ and $y_\tau$) and three scalar quartic couplings ($\lambda_2$ and $\lambda_3$). The prior intervals for these parameters are given in Table (\ref{tab:param:range})

\begin{table}[!t]
    \setlength\tabcolsep{40pt}
    \centering
    \begin{tabular}{c c}
    \toprule
    Parameter     & Range  \\
    \toprule
    $\lambda_2$   & $[0, 4 \pi]$ \\
    $\lambda_3$   & $[-4 \pi, 4 \pi]$ \\
    $y_\ell; \ell=e, \mu,\tau$         & $[10^{-5}, 4 \pi]$ \\
    $m_{H^\pm}$   & $[50, 1000]~{\rm GeV}$ \\
    $m_{N}$       & $[10~{\rm GeV}, m_{H^\pm}[$ \\
    \toprule
    \end{tabular}
    \caption{The independent parameters of the model and their range.}
    \label{tab:param:range}
\end{table}

We close this section by noting that the interaction \eqref{eq:int:1} can be obtained from a UV complete model. We list the following possibilities\footnote{We leave this for a future study \cite{Jueid:2020xx}.}:
\begin{itemize}
\item From the embedding of the SM into a $SU(5)$ gauge group. The matter fields are chosen to belong to the ${\bf{10}}_\alpha$ and $\bar{{\bf{5}}}_\alpha$ representations. The charged scalar singlet belongs to the ${\bf{10}}_H$ representation and the right-handed Majorana fermion is incorporated by adding a singlet to the minimal setup of $SU(5)$. The interaction Lagrangian invariant under $SU(5)$ is given by
\begin{eqnarray}
	\mathcal{L}_{\rm int} = y_{\alpha\beta} \overline{\bf{10}}_\alpha \otimes {\bf{10}}_H \otimes {\bf{1}}_{N_\beta} \supset y_{\alpha \beta} \ell^T_{R\alpha} C N_\beta S^+.
\end{eqnarray}
\item From the embedding of the SM into a flipped $SU(5)\otimes U(1)_X$ grand-unified theory. In this case, the scalar singlet belongs to the singlet representation and the interaction in \eqref{eq:int:1} can be obtained from the following effective interaction:
\begin{eqnarray}
\mathcal{L} = \frac{h_{\alpha \beta}}{\Lambda} \overline{{\bf{10}}}_\alpha \otimes \bar{{\bf{1}}}_\beta \otimes {\bf{10}}_{H} \otimes {\bf{1}}_{S} + h.c. \supset \frac{h_{\alpha\beta} \langle {\bf{10}}_H \rangle}{\Lambda} N^T C \ell_R S^-
\end{eqnarray}
\end{itemize}

\section{Theoretical and experimental constraints}
\label{sec:constraints}
This model is subject to various theoretical and experimental constraints (the effects of these constraints is summarized in Figs. \ref{fig:bounds:1}-\ref{fig:bounds:2}) :

\begin{description}
\item[Theoretical constraints:] First we require that the parameters of the scalar potential of the model satisfy a number of constraints.

\begin{itemize}
\item {\it Boundedness-from-below:} The quartic scalar couplings of the scalar potential should satisfy the following conditions \cite{Branco:2011iw}: $\lambda_{1, 2} > 0,~ \lambda_3+2\sqrt{\lambda_1 \lambda_2} > 0$.
\item {\it Perturbativity:} The scalar quartic couplings should obey to the condition; $\lambda_i \leq 4 \pi$. 
\item {\it Perturbative unitarity:} Variety of scattering processes should maintain perturbativity at high energies \cite{Kanemura:1993hm, Akeroyd:2000wc}. We note that the conditions of the perturbative unitarity in our model can be obtained from those in the Inert Higgs Doublet Model (IHDM) by simply setting $\lambda_4 = \lambda_5 = 0$ \cite{Arhrib:2015hoa}.
\item {\it False vacuum:} Should be avoided by simply setting the following condition \cite{Ginzburg:2010wa}
$$
\frac{m_{11}^{2}}{\sqrt{\lambda_{1}}}\geq-\frac{2 m_{H^\pm}^{2} - \lambda_3 \upsilon^2}{2 \sqrt{\lambda_{2}}}
$$
\end{itemize}

\item[Higgs signal strength measurements:] In our model, the only modification to the Higgs boson production and decay comes from the effect of the charged singlet scalar on the one-loop induced $\Gamma(H\to\gamma\gamma)$ \cite{Swiezewska:2012eh, Jueid:2020rek}. The partial decay width of $H\to\gamma\gamma$ is given by \cite{Swiezewska:2012eh}
\begin{eqnarray}
 \Gamma(H\to\gamma\gamma) = \frac{G_F \alpha^2 m_H^3}{128 \sqrt{2} \pi^3} \bigg|\sum_{f=\tau,b,t} N_{c,f} Q_i^2 A_{1/2}(\tau_f^2) + A_{1}(\tau_W^2) + \frac{\lambda_3 v^2}{2 m_{H^\pm}^2} A_0(\tau_{H^\pm}^2) \bigg|^2, \hspace{0.7cm}
\end{eqnarray}
with $\tau_i = 2 m_i/m_H$, $N_{cf}$ is the number of colors for the fermion $f$, and $A_{i}(\tau_i)$ are the one-loop form factors whose expressions can be found in {\it e.g.,} \cite{Swiezewska:2012eh}. In this work, we use the combined measurement by \textsc{Atlas} and \textsc{Cms}  \cite{Khachatryan:2016vau}:
$$
|\kappa_\gamma| = \sqrt{\Gamma(H\to\gamma\gamma)/\Gamma(H\to\gamma\gamma)_\mathrm{SM}} = 0.87^{+0.14}_{-0.09}
$$
We require that $\kappa_\gamma$ in our model to be within the $2\sigma$-level of the corresponding experimental measurement.

\item[Higgs invisible decay:] Strong constraints on the Yukawa coupling $y_N$ comes mainly from the limits on the Higgs invisible decay branching fractions; $\mathrm{BR}(H\to N_R N_R) \equiv \textrm{B}_{\mathrm{inv}}$ which becomes effective for $m_{N} < m_H/2$. In our model, the invisible decay of the SM Higgs boson is one-loop induced. We have used \texttt{FeynArts}, \texttt{FormCalc} and \texttt{LoopTools} \cite{Hahn:2000jm, Hahn:2000kx} to compute ${\rm B}_{\rm inv}$. Assuming that $m_\ell \simeq 0$, we get

\begin{eqnarray}
y_N < \bigg(\frac{2048 \pi^5 \Gamma_{H}^{\mathrm{SM}}}{\beta_N^{3/2} m_H \lambda_3^2 v^2 m_N^2 |C_0 + C_2|^2 \left(\frac{1}{B_\mathrm{upper}} - 1\right)}\bigg)^{1/4},
\end{eqnarray}

with $\beta_N = (1 - 4 m_N^2/m_H^2)$, and $C_{0,2} \equiv C_{0,2}(m_N^2,m_H^2,m_N^2,0,m_{H^\pm}^2,m_{H^\pm}^2)$ are the three-point Passarino-Veltman functions \cite{Passarino:1978jh}.

\begin{figure}[!t]
\includegraphics[width=0.48\linewidth]{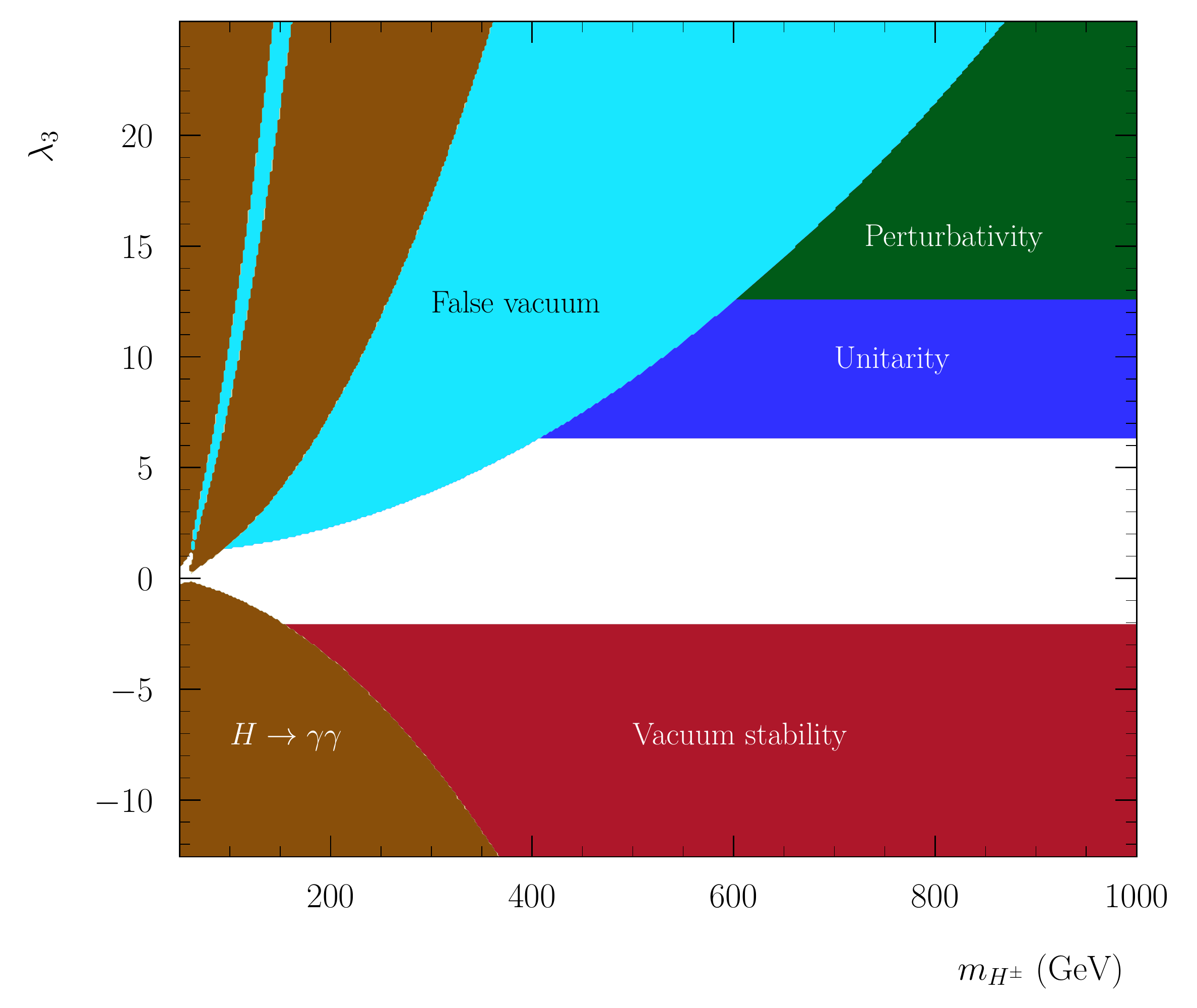}
\hfill
\includegraphics[width=0.48\linewidth]{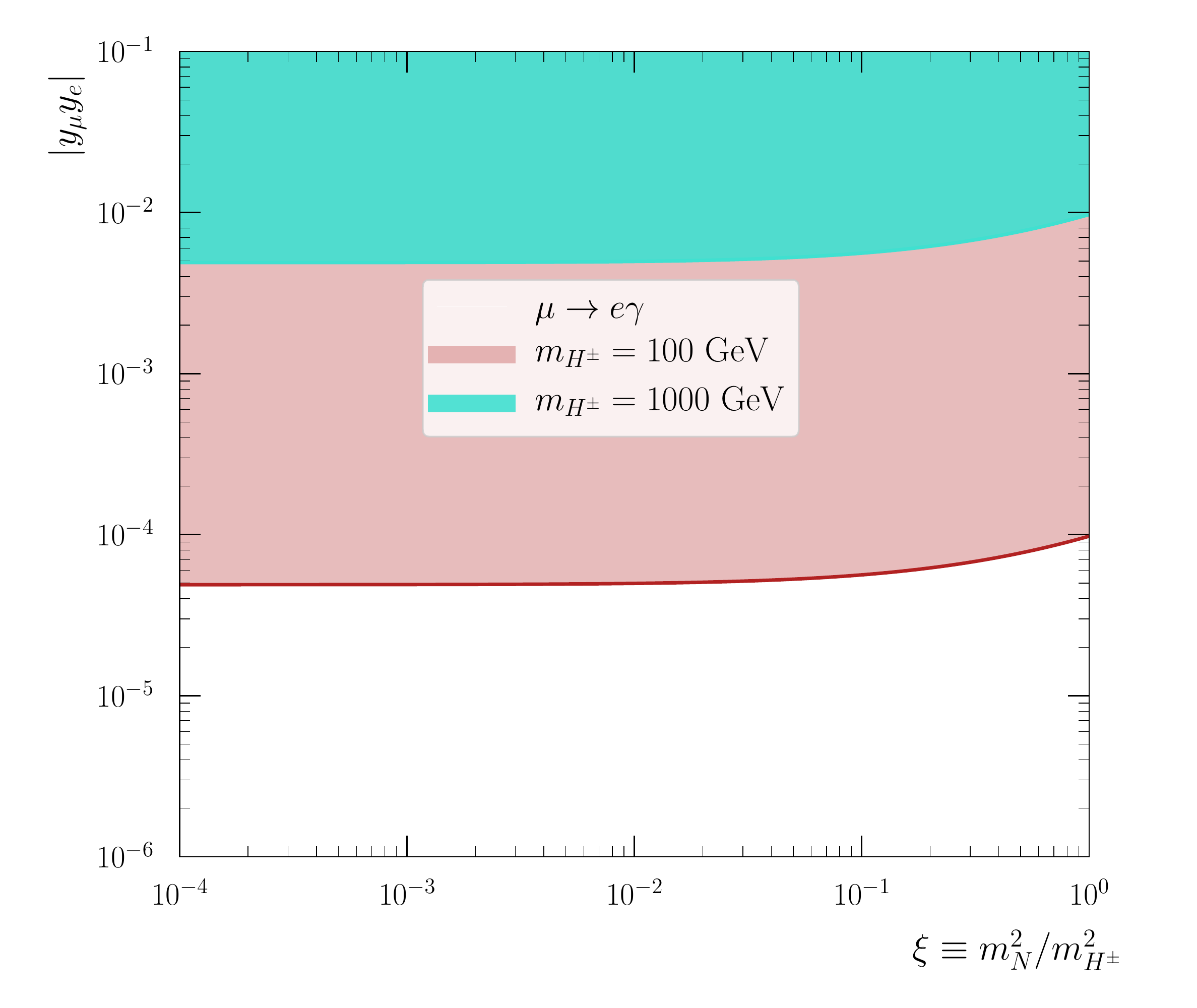}
\caption{\emph{Left:} Summary of the impact of theoretical and experimental constraints on the model parameter space projected on the ($m_{H^\pm}, \lambda_3$) plane. \emph{Right:} Bounds on $|y_e y_\mu|$ from the \textsc{MeG} searches of $\mu\to e\gamma$ for $m_{H^\pm} = 100~{\rm GeV}$ (red shaded area) and $m_{H^\pm} = 1000~{\rm GeV}$ (green shaded area).}
\label{fig:bounds:1}
\end{figure}

\begin{figure}[!t]
\includegraphics[width=0.48\linewidth]{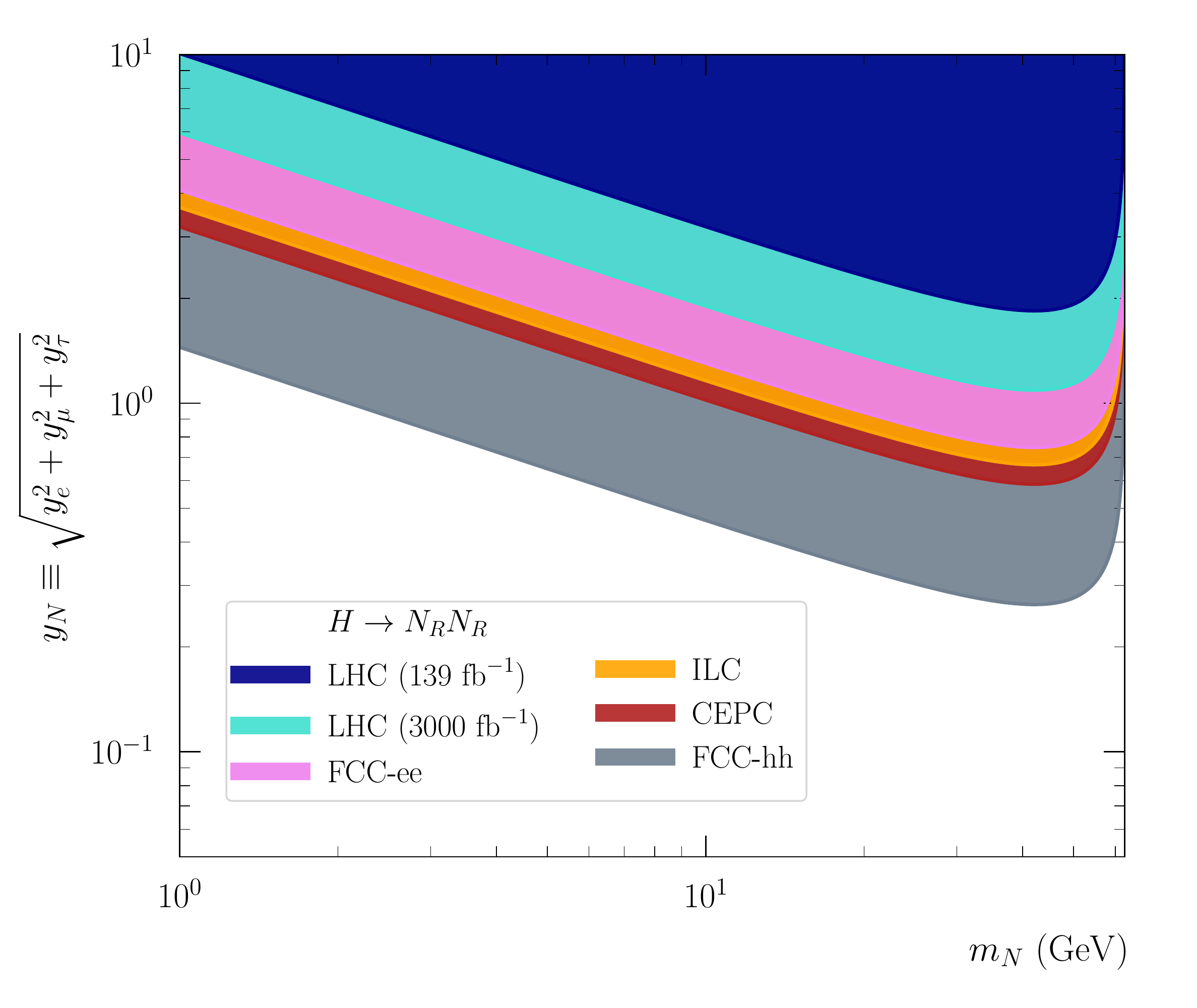}
\hfill
\includegraphics[width=0.48\linewidth]{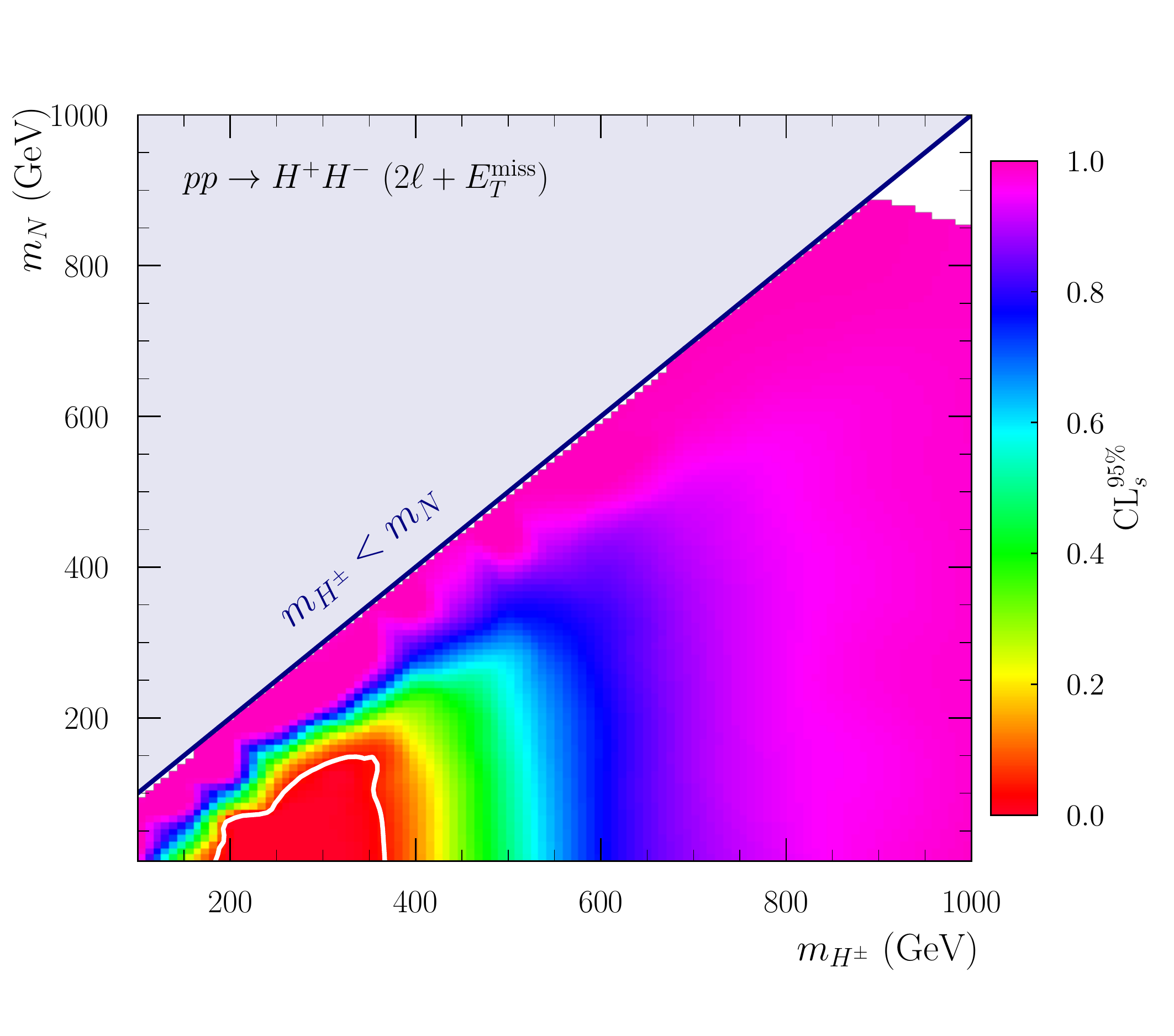}
\caption{\emph{Left:} Present and expected bounds on $y_N$ from $H \to N_R N_R$ decay for $\lambda_3 = 1$ and $m_{H^\pm} = 100~{\rm GeV}$. In  the same panel, we show the bounds on $y_N$ from the LHC \cite{Sirunyan:2018owy}, HL-LHC \cite{CMS-PAS-FTR-16-002}, FCC-ee \cite{Cerri:2016bew}, ILC \cite{Asner:2013psa}, CEPC \cite{CEPCStudyGroup:2018ghi} and FCC-hh \cite{Selvaggi:2018obt}. \emph{Right:} The LHC exclusions from slepton searches at $\sqrt{s} = 13~{\rm TeV}$ and integrated luminosity of $139~{\rm fb}^{-1}$. The color code corresponds to the ${\rm CL}_s$ value projected on the $(m_{H^\pm}, m_N)$ plane. The white solid line corresponds to ${\rm CL}_s = 5 \times 10^{-2}$ and the region bounded by this line is excluded at the $95\%$ CL.}
\label{fig:bounds:2}
\end{figure}

\item[Charged Lepton flavor violating (CLFV) decays:] The interaction in eqn \eqref{eq:int:1} breaks the generation lepton number by one unit; {\it i.e.,} $\Delta N_\ell \neq 0$ with $\ell=e, \mu, \tau$. Therefore, this model implies CLFV induced at the one-loop order. We use the results from the \textsc{Meg} \cite{Adam:2013mnn} and \textsc{BaBar} \cite{Aubert:2009ag} experiments to derive bounds on the product of the Yukawa couplings $|y_{\ell_\alpha} y_{\ell_\beta}|$: 
\begin{eqnarray}
	|y_e y_\mu| \simeq 0.92 \times 10^{-2} |y_\tau y_\mu| \simeq 10^{-2} |y_e y_\tau| < \bigg(\frac{2.855 \times10^{-5}}{\mathrm{GeV}} \bigg)^2 \frac{m_{H^\pm}^2}{|\mathcal{F}(m_N^2/m_{H^\pm}^2)|}, \hspace{0.5cm}
\end{eqnarray}

where $\mathcal{F}(X)$ is the one-loop form factor whose expression can be found in {\it e.g.,} \cite{Ahriche:2017iar}. In this study, we choose a scenario defined as $y_e \simeq \mathcal{O}(1) \gg y_\tau \gg y_\mu$ while the implications of different hierarchies will be dedicated for a future study. \\

\item[LEP constraints:] The model can be subject to constraints from searches of new physics at LEP. For instance, using the void results of chargino searches at center-of-mass energies of $\sqrt{s}=183$-$209$ GeV carried by the OPAL collaboration, we can derive bounds on $m_{H^\pm}$ for some choices of $y_e \simeq y_N$. We got the following bound
$$
m_{H^\pm} > 90~{\rm GeV}, \qquad {\rm for}~y_e = 2.
$$
More details can be found in {\it e.g.,} \cite{Ahriche:2018ger}. 

\item[LHC constraints:]
This model can be constrained from searches of sleptons/charginos at the LHC. The reason is that the interaction in \eqref{eq:int:2} leads to the production of singlet charged scalars through $q\bar{q}$ annihilation with the exchange of $\gamma^*/Z$. The decays of the charged scalars lead to a final state formed by two charged leptons and large missing energy. Due to our choice of $y_\ell$, this final state is electron enriched. In order to study these constraints, we employ the most recent search of supersymmetric particles at $\sqrt{s} = 13~{\rm TeV}$ and $\mathcal{L} = 139~{\rm fb}^{-1}$. For this purpose, we use the \texttt{MadAnalysis} 5 framework \cite{Conte:2012fm,Conte:2014zja,Conte:2018vmg, Fuks:2021wpe, Araz:2020dlf} to obtain the constraints from the reinterpretation of this ATLAS search. More details about the implementation can be found in ref. \cite{Araz:2020dlf} and detailed results within this model can be found in ref. \cite{Jueid:2020yfj}. 
\end{description}

\section{Dark Matter correlations}
\label{sec:DM}

The relic density of $N_R$ gets several contributions:
\begin{itemize}
\item Annihilation channels which dominate for $\Delta \equiv (m_{H^\pm} - m_N)/m_N \geqq 0.1$: 
      \begin{enumerate}
      \item[{\it (i)}] $N_R N_R \to \ell^+ \ell^-$ through the exchange of the charged singlet scalar $H^\pm$ in $t$- and $u$-channels.
      \item[{\it (ii)}] $N_R N_R \to f \bar{f}, WW, ZZ, \gamma\gamma~{\rm and}~HH$ through the exchange of the SM Higgs boson in $s$-channel.
      \end{enumerate}
\item Co-annihilation channels which are significantly important for $\Delta < 0.1$:
\begin{enumerate}
\item[{\it (i)}]	 $H^\pm H^\mp \to \ell^\pm \ell^\mp, q\bar{q}, HH, ZZ, W^\pm W^\mp, Z H, t\bar{t}$
\item[{\it (ii)}] $N_R H^\pm \to \ell^\pm H, W^\pm \nu, \ell^\pm Z, \ell^\pm \gamma$
\end{enumerate}
\end{itemize}

Note that the contribution of the annihilation through the exchange of the SM Higgs boson in the $s$-channel is always sub-leading ($\leq 0.1\%$). Co-annihilation channels dominate over annihilation if $\Delta < 0.1$ and for some values of the couplings {\it e.g,} $\lambda_3$. All the annihilation and the co-annihilation channels are included in our analysis.

\begin{figure}[!t]
    \centering
    \includegraphics[width=0.89\textwidth]{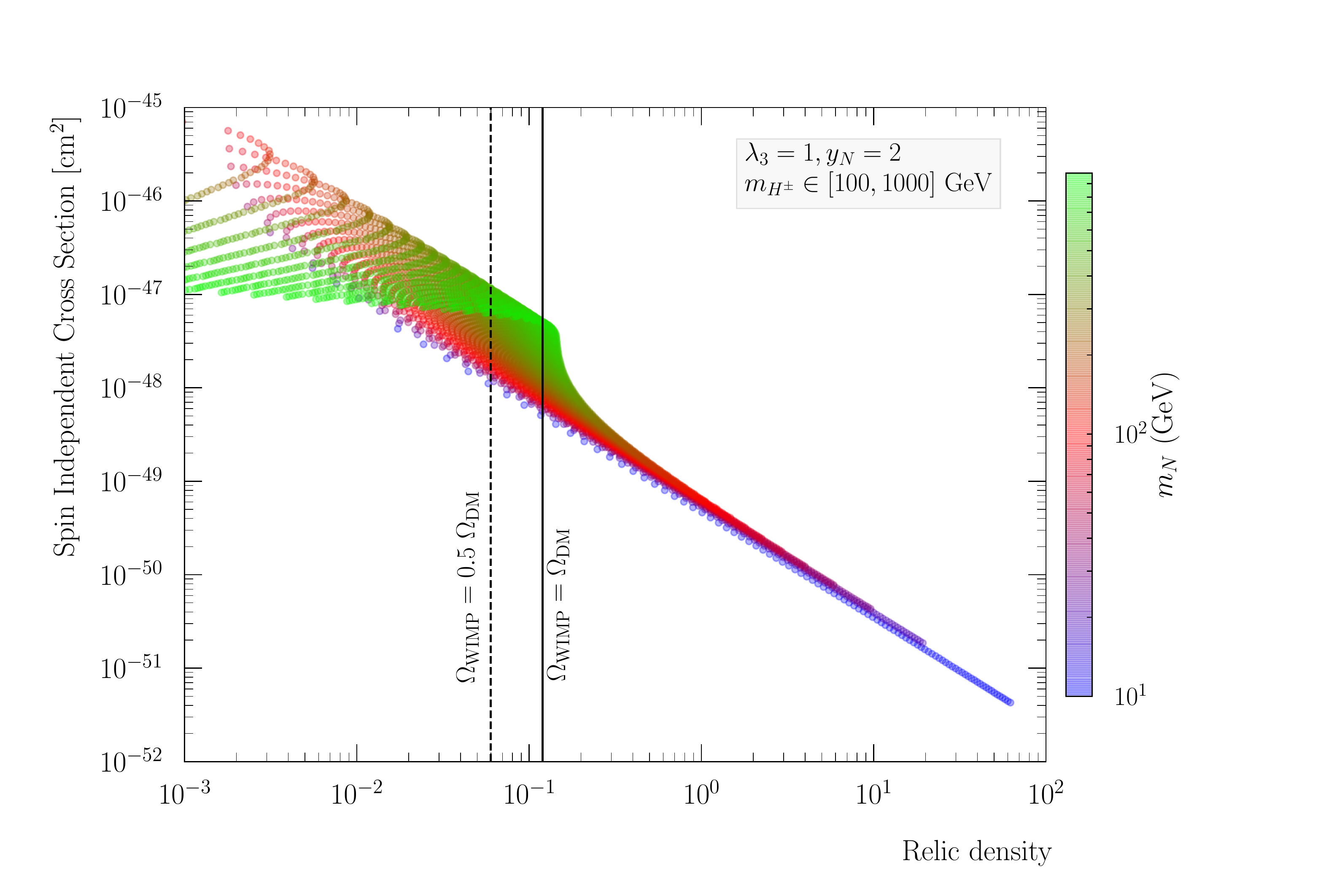}
    \caption{The correlation between the relic abundance $\Omega_{N_R} h^2$ and the spin-independent direct detection cross section $\sigma_{\rm SI}$. The color code corresponds to the values of the DM mass $m_N$.}
    \label{fig:Correlations}
\end{figure}

The  scattering of $N_R$ off nuclei occurs at the one-loop level  through the exchange of the SM Higgs boson. Due to the nature of $N_R$, the elastic  scattering is dominated by the spin-independent contribution which is given by 
\begin{eqnarray}
 \sigma_{\rm SI} = \frac{4}{\pi} \mu_{\mathcal{N}}^2 \bigg( Z \cdot f_p + (A - Z) \cdot f_n \bigg)^2,
\end{eqnarray}
with $f_{p,n}$ are the nucleons ($p/n$) spin-independent form factors, and $\mu_{\mathcal{N}} \equiv m_N m_{\mathcal{N}}/(m_N + m_{\mathcal{N}})$ is the reduced mass of the DM-$\mathcal{N}$ system. The scattering cross section depends on the square of the effective $\tilde{y}_{HNN}$ coupling \cite{Ahriche:2017iar}

\begin{eqnarray}
\tilde{y}_{HNN} \simeq -\frac{\lambda_3 v |y_N|^2}{16 \pi m_N} \bigg[1 - \left(1 - r_N^{-2} \right) \log{\left(1 - r_N^2 \right)} \bigg], 
\end{eqnarray}
with $r_N = m_{N}/m_{H^\pm}$. The spin-independent cross section scales as $\lambda_3^2 |y_N|^4/m_N^2$ times factors that depend on the mass splitting $\Delta_{H^\pm} = m_{H^\pm} - m_N$. We have used \texttt{MadDM} version 3.0 \cite{Ambrogi:2018jqj} to compute both the relic abundance ($\Omega_{N_R} h^2$) as well as the spin-independent Nucleus-DM elastic cross section ($\sigma_{\rm SI}$). \\

It is found that the condition $y_N > 0.1$ is required to avoid the over-abundance of $N_R$ particles (regardless of the choice of $\lambda_3$). For $\Delta < 0.1$, it is found that the criterion on $y_N$ is extremely relaxed as the correct relic abundance is guaranteed by the choice of $\lambda_3$ of order $\mathcal{O}(1)$. On the other hand, $y_N$ of order $\mathcal{O}(0.1)$ yields an extremely small $\sigma_{\rm SI}$ close to the neutrino-floor. In Fig. (\ref{fig:Correlations}), we display the correlation between the relic abundance and the spin-independent scattering cross section $\sigma_{\rm SI}$. We can see that there is a strong anti-correlation between these two observables within our model. The parameter space regions where a correct relic density is produced  $\Omega h^2 \simeq \Omega_\mathrm{Planck} h^2$ correspond to masses in the range $ 200$-$750$ GeV for which the corresponding spin-independent cross section is about $\simeq 10^{-48} ~\mathrm{cm}^2$ which is below the \textsc{Xenon1T} bound \cite{Aprile:2018dbl}. In the following section, we choose the following benchmark scenarios:

\begin{equation}
100 \leqslant m_{H^\pm}/\mathrm{GeV} \leqslant 1000, \quad 10 \leqslant m_N/\mathrm{GeV} < m_{H^\pm},~\lambda_3 =1,~y_N = 2.
\label{eq:BPs}
\end{equation}

\section{Phenomenology at the International Linear Collider}
\label{sec:ILC}

\begin{figure}
\centering
\includegraphics[width=1.0\linewidth]{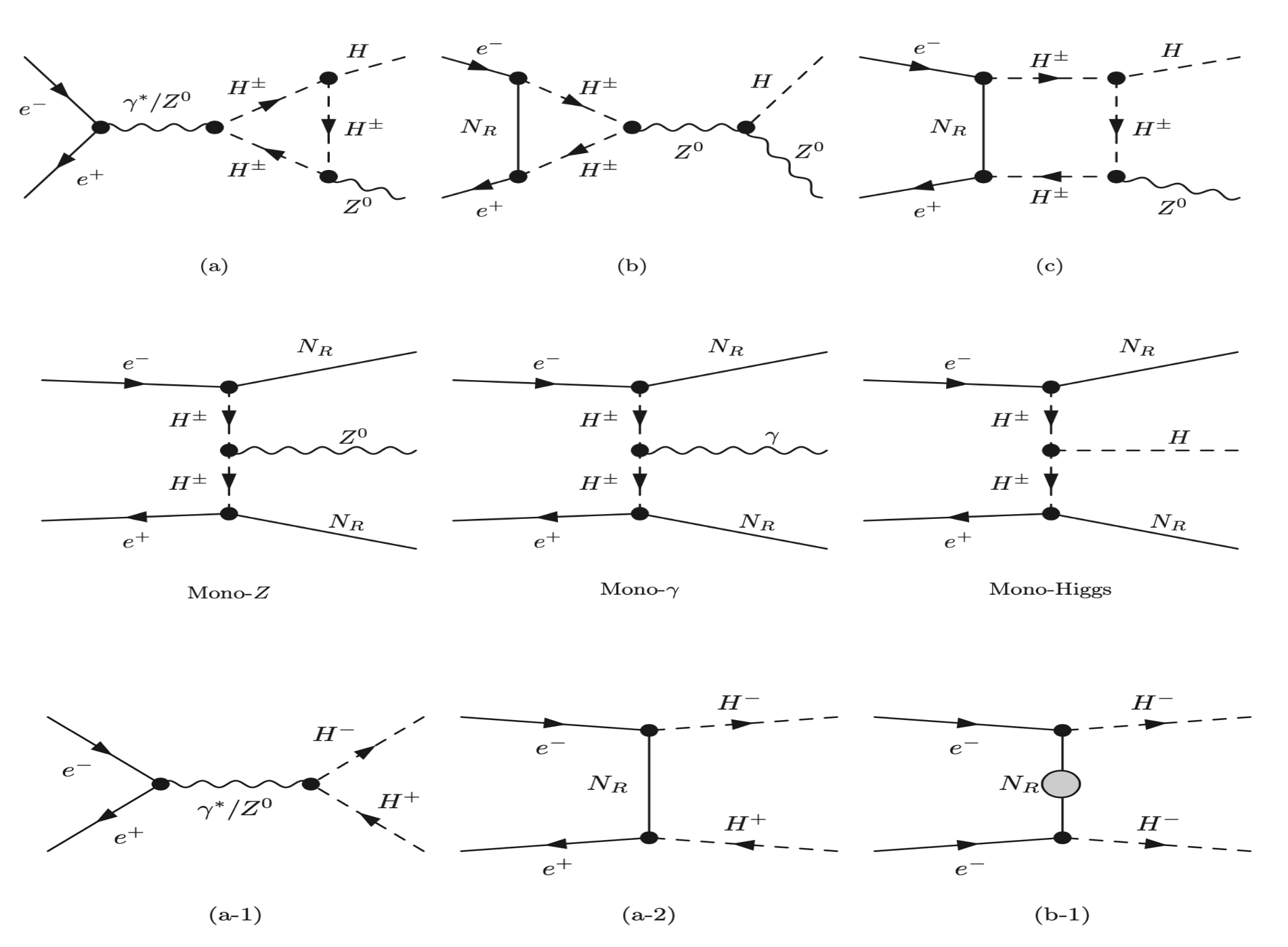}
\caption{Example of processes which  can occur in high energy $ee$ colliders. {\it Top panels:} Higgs-strahlung ($e^+ e^- \to HZ^0$) where our model contributes at the one-loop order. {\it Middle panels:} Mono-$X$ processes with $X = Z^0, \gamma, H$. {\it Bottom panels:} Charged Higgs pair production.}
\label{fig:ILC:processes}
\end{figure}

\subsection{Introduction}
This model can be tested at the International Linear Collider (ILC) through several processes. The ILC is expected to run at center-of-mass energies of $250, 350, 500$  and $1000$ GeV \cite{Baer:2013cma}. In Fig. (\ref{fig:ILC:processes}), we list the generic processes which can be used to test our model: 

\begin{itemize}
\item Contribution to $\sigma(e^+ e^- \to HZ)$ at the one-loop order -- top panels of Fig. (\ref{fig:ILC:processes}) --. The rate of $HZ$ production at  lepton colliders at the one-loop gets contributions from: \emph{(i)} triangle corrections to the initial $e^+ e^- Z$ and the final $\gamma^*/Z Z H$ vertex with the exchange of $N_R$ and $H^\pm$ and \emph{(ii)} box corrections with $N_R$ and $H^\pm$ running in the loop. Note that box corrections may involve non-trivial effects on the differential distributions; bump-like structures in {\it e.g.,} the transverse momentum of the SM Higgs boson.
\item Mono-$X$ processes, {\it i.e.,} $e^+ e^- \to N_R N_R X; X=\gamma,Z^0,H$ -- middle panels of Fig. (\ref{fig:ILC:processes}) --. \emph{(i)}: $e^+ e^- \to N N Z$ which yields a signature dubbed mono-$Z$ whose rate is proportional to $y_e^4$, \emph{(ii)}: $e^+ e^- \to N N \gamma$ which yields mono-photon and whose rate is proportional to $y_e^4$,  \emph{(iii)}: $e^+ e^- \to N N H$ which yields a variety of final states depending on the decay products of the SM Higgs boson with a rate increasing with $y_e^4 \lambda_3^2$.
\item Charged scalar pair production which includes opposite-sign ($e^+ e^- \to H^+ H^-$) and same-sign ($e^- e^- \to H^- H^-$) processes --bottom panels of Fig. (\ref{fig:ILC:processes}) --. The charged Higgs pair production in the electron-electron option at the ILC is more interesting than the opposite sign $H^+ H^-$ due to the fact that: \emph{(i)} the signal process gets a $m_N^2$ enhancement, and \emph{(ii)} the associated backgrounds have very small rates.
\end{itemize}

\subsection{${\rm Higgs} + E_{T}^{\rm miss}$ at the ILC}

We study the the Higgs + $E_{T}^{\rm miss}$ process at two center-of-mass energies; $\sqrt{s} = 500~{\rm GeV}$ and $\sqrt{s} = 1000~{\rm GeV}$. To optimise the event rate at the ILC, we consider the Higgs boson decaying into $b\bar{b}$. This will lead to a final state comprising two $b$-tagged jets and large missing energy. The major backgrounds we need to deal with include $H(\to b\bar{b})Z(\to\nu\bar{\nu})$, $H(\to b\bar{b})\nu_e \bar{\nu}_e$, $W^+ W^-$, $ZZ$, and $t\bar{t}$. For $[P_{e^-}, P_{e^+}] = [+0.8, -0.3]$ and $\sqrt{s} = 500~{\rm GeV}$, the total cross section for the background processes reaches about $2~{\rm pb}$. We have used \texttt{Madgraph5\_aMC@NLO} version 2.6.7 \cite{Alwall:2014hca} to generate Monte Carlo (MC) samples for both the signal and the backgrounds. The decays of resonant states, parton showering of coloured particles and hadronisation were handled using \texttt{Pythia8} version 8.2.44 \cite{Sjostrand:2014zea}. Fast detector simulation has been performed using \texttt{Delphes} 3.4.2 \cite{deFavereau:2013fsa}. The clustering of jets is done according to the anti-$k_T$ algorithm \cite{Cacciari:2008gp} with a jet radius $R=0.4$ using \texttt{FastJet} version 3.3.1 \cite{Cacciari:2011ma}. Events are selected if they contain exactly two $b$-tagged jets with $p_\perp > 30$ GeV and $|\eta| < 2.5$. Furthermore, events are vetoed if they contain one isolated charged lepton (electron or muon) with $p_\perp^\ell > 15$ GeV and $|\eta^\ell| < 2.5$. The two $b$-tagged jets are used to reconstruct a Higgs boson {\it candidate}. First, we require that the transverse momentum of the $b\bar{b}$ pair is larger than $50~{\rm GeV}$; $p_\perp^{b\bar{b}} > 50~{\rm GeV}$. In this fiducial phase space region, the four-momentum of the 
invisible system can be determined from the center-of-mass energy and the four-momentum of the reconstructed SM Higgs boson; we get
\begin{eqnarray}
M_{\mathrm{inv}}^2 = s - 2 \sqrt{s} E_{b\bar{b}} + M_{b\bar{b}}^2,
\end{eqnarray} 
where $s$ is the center-of-mass energy, and  $E_{b\bar{b}}$ ($M_{b\bar{b}}$) is the energy (the invariant mass) of the $b\bar{b}$ system. The signal region is defined as 
\begin{eqnarray}
|M_{b\bar{b}} - m_H| < 10~{\rm GeV}; \quad m_H = 125~{\rm GeV} \\ 
200~{\rm GeV} < M_{\rm inv} < 800~{\rm GeV}.
\end{eqnarray}
After the full selection, the acceptance times the efficiency of the signal varies in the range $2$-$3\%$ while the background rejection is about $200$. We note, finally, that the selection efficiency does not depend on the polarisation of the initial beams.

\subsection{Same-sign charged scalar pair production}
We study the same-sign charged scalar pair production at $\sqrt{s}_{ee} = 1000~{\rm GeV}$. The cross section for the signal behaves as \cite{Aoki:2010tf}
\begin{eqnarray*}
\sigma_{e^- e^- \to H^- H^-} &=& \int_{t_{\rm min}}^{t_{\rm max}} \frac{{\rm d}t}{128 \pi s} \bigg|y_e^2 m_N \bigg(\frac{1}{t - m_N^2} + \frac{1}{y - m_N^2}\bigg)\bigg|^2 \times \frac{\Gamma_{H^- \to \ell^-_\alpha N_R} \Gamma_{H^- \to \ell^-_\beta N_R}}{\Gamma_{H^-}^2}, \\
&\propto & m_N^2 y_e^4 \frac{y_{\ell_\alpha}^2 y_{\ell_\beta}^2}{(y_e^2 + y_\mu^2 + y_\tau^2)^2},
\end{eqnarray*}
where we have used the narrow-width approximation in the first line and neglected mass effects in the second line taking the leading terms in both the production and the decay stages. Due to our choice of $y_\ell$ from CLFV constraints, we consider that both of the charged singlet scalars decay into $e^- N_R$. The final state is composed of two same-sign electrons and large missing energy. The main backgrounds for this process are $W^- W^- \nu_e \nu_e$ and $Z e^- e^-$ whose cross sections are:
$$
\sigma_{e^- e^- \to W^- W^- \nu_e \nu_e} \times {\rm BR}_{W^- \to e^- \bar{\nu}_e}^2 \simeq 0.22~{\rm fb} \quad {\rm and} \quad \sigma_{Z e^- e^-} \times {\rm BR}_{Z \to \nu \bar{\nu}} = 19.62~{\rm fb}.
$$ 
This has to be compared with the cross sections for the signal process. For example, we have $\sigma_{e^- e^- \to H^- H^-}  = 5.46~(32.94)$ pb for $m_{H^\pm} = 250$ GeV and $m_N = 30~(180)$ GeV. This already implies a signal-to-background ratio ($n_S/n_B$) that varies between $200$ and $1750$. Given that large signal-to-background ratios, we only apply some minor selections on the decay products of the charged singlet scalar at the detector level; \emph{i.e.,} we select events if they contain exactly two same-sign electrons with
\begin{eqnarray}
p_\perp^e > 20~{\rm GeV},	\quad |\eta^e| < 2.5
\end{eqnarray}

\subsection{Results}

\begin{figure}[!t]
    \centering
    \includegraphics[width=1.01\columnwidth]{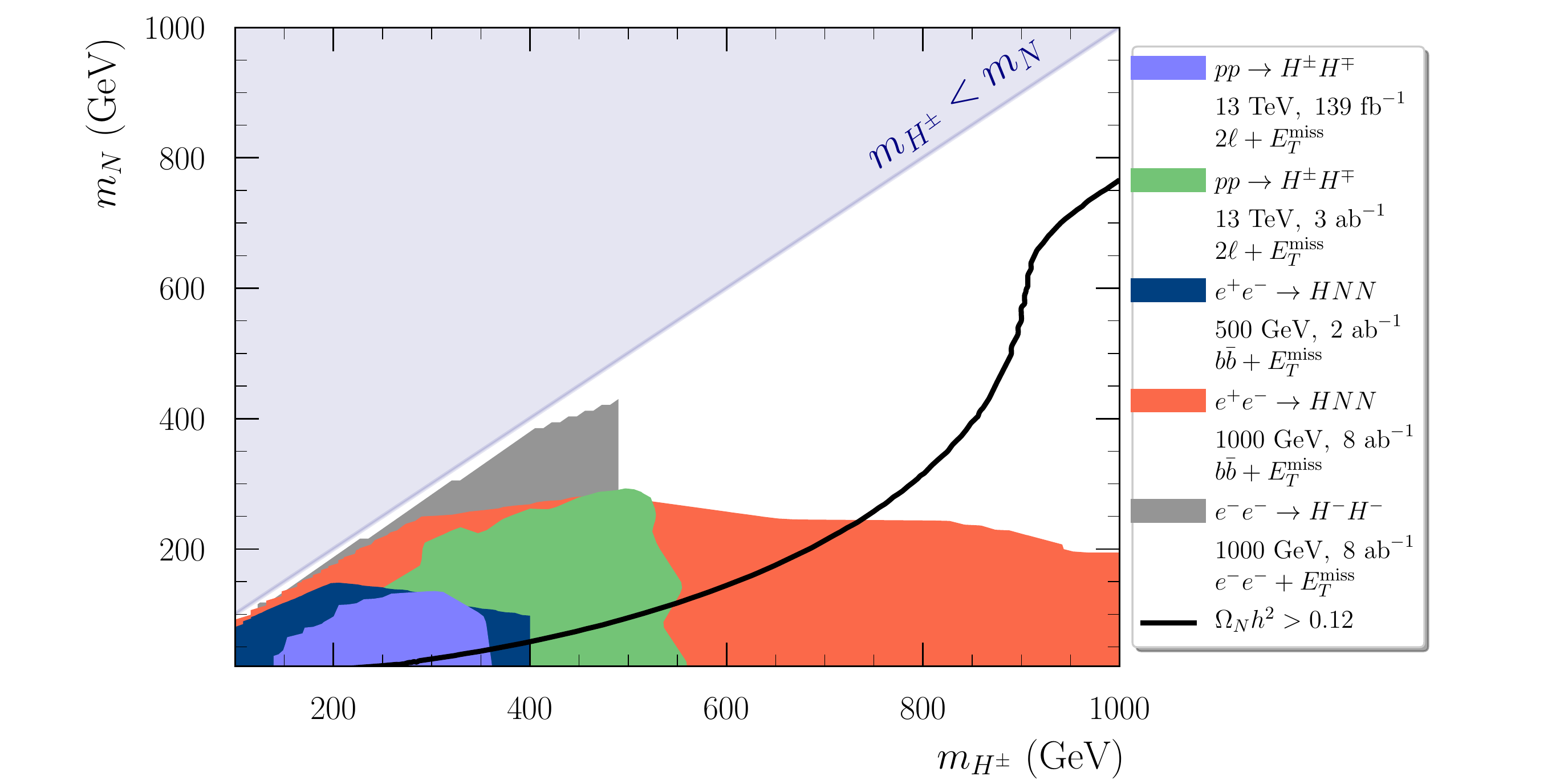}
    \caption{The future expected exclusions projected on the $(m_{H^\pm}, m_N)$ plane for $y_N = 2$ and $\lambda_3 = 1$. The limits correspond to the future HL-LHC at $\sqrt{s} = 13~{\rm TeV}$ (green), mono-Higgs process at $\sqrt{s} = 500~{\rm GeV}$ (blue) and $\sqrt{s} = 1000~{\rm GeV}$ (red) and the same-sign singlet pair production at $\sqrt{s} = 1000~{\rm GeV}$ (gray).}
    \label{fig:sensitivity-ILCLHC}
\end{figure}

To obtain the expected exclusions on the parameter space of our model from mono-Higgs process and the production of the same-sign charged singlet scalar at the ILC, we compute the CL$_s$ values \cite{Read:2002hq}:
\begin{eqnarray}
{\rm CL}_{s} = \max\left(0, 1 - \frac{p_{s+b}}{p_b}\right),
\end{eqnarray}
with $p_b$, and $p_{b+s}$ are the background and the signal-plus-background probabilities. In our analysis, we used $5 \times 10^5$ toy experiments to estimate these probabilities. For the calculation of the ${\rm CL}_s$ values, we assume a flat $10\%$ uncertainty on the background yields.
A  point in the parameter space of the model is excluded at the $95\%$ CL if ${\rm CL}_s < 0.05$. In  Fig. \ref{fig:sensitivity-ILCLHC}, we show the future exclusions projected on the mass of the charged singlet scalar $m_{H^\pm}$ and the mass of the Majorana dark matter $m_N$ for $y_N = 2$ and $\lambda_3 = 1$. We draw the following conclusions: {\it (i)} The pair production of charged singlet scalars at hadron colliders cannot probe the region of the parameter with $\Delta \equiv m_{H^\pm} - m_N < 50~{\rm GeV}$. {\it (ii)} The mono-Higgs process can be used to probe dark-matter masses of up to $200~{\rm GeV}$ regardless of the values of $\Delta$ and {\it (iii)} the same-sign singlet charged scalar pair production can probe the whole parameter space allowed by the center-of-mass energy of the ILC, {\it i.e.,} $m_{H^\pm} \leq \sqrt{s}/2$. 

\section{Conclusions}
\label{sec:conclusion}
In this work, we have proposed a very minimal extension of the SM to account for a dark-matter candidate and studied its phenomenology at future Linear Colliders. The theoretical setup consists of extending the SM with two $Z_2$-odd, $SU(2)_L$ gauge singlets: a Majorana fermion $N_R$ and charged scalar $S$. First, we have studied theoretical constraints and constraints from all the available searches and measurements: Higgs precision data, searches of CLFV decays, searches of Higgs invisible decays, searches of charginos at LEP and searches of sleptons at the LHC. Then, we have shown that there is a strong anti-correlation between the relic abundance and the spin-independent nucleus-$N_R$ scattering cross section. On the other hand, for parameters allowed by the theoretical constraints, we found that the spin-independent cross section is neither excluded nor close to the neutrino floor. Finally, we studied the prospects at future lepton colliders where we have chosen two processes; mono-Higgs production with the SM Higgs boson decaying into $b\bar{b}$ and the pair production of same-sign charged singlet scalars. 

\section*{Acknowledgments} 
The work of AJ is supported by the National Research Foundation of Korea, Grant No. NRF-2019R1A2C1009419.  

\bibliographystyle{JHEP}
\bibliography{main.bib}

\providecommand{\href}[2]{#2}\begingroup\raggedright\begin{thebibliography}{10}

\bibitem{Aprile:2018dbl}
{\bf XENON} Collaboration, E.~Aprile et~al., {\it {Dark Matter Search Results
  from a One Ton-Year Exposure of XENON1T}},  {\em Phys. Rev. Lett.} {\bf 121}
  (2018), no.~11 111302, [\href{http://arxiv.org/abs/1805.12562}{{\tt
  arXiv:1805.12562}}].

\bibitem{Cui:2017nnn}
{\bf PandaX-II} Collaboration, X.~Cui et~al., {\it {Dark Matter Results From
  54-Ton-Day Exposure of PandaX-II Experiment}},  {\em Phys. Rev. Lett.} {\bf
  119} (2017), no.~18 181302, [\href{http://arxiv.org/abs/1708.06917}{{\tt
  arXiv:1708.06917}}].

\bibitem{Adriani:2010rc}
{\bf PAMELA} Collaboration, O.~Adriani et~al., {\it {PAMELA results on the
  cosmic-ray antiproton flux from 60 MeV to 180 GeV in kinetic energy}},  {\em
  Phys. Rev. Lett.} {\bf 105} (2010) 121101,
  [\href{http://arxiv.org/abs/1007.0821}{{\tt arXiv:1007.0821}}].

\bibitem{Aguilar:2013qda}
{\bf AMS} Collaboration, M.~Aguilar et~al., {\it {First Result from the Alpha
  Magnetic Spectrometer on the International Space Station: Precision
  Measurement of the Positron Fraction in Primary Cosmic Rays of 0.5–350
  GeV}},  {\em Phys. Rev. Lett.} {\bf 110} (2013) 141102.

\bibitem{Ahnen:2016qkx}
{\bf MAGIC, Fermi-LAT} Collaboration, M.~Ahnen et~al., {\it {Limits to Dark
  Matter Annihilation Cross-Section from a Combined Analysis of MAGIC and
  Fermi-LAT Observations of Dwarf Satellite Galaxies}},  {\em JCAP} {\bf 02}
  (2016) 039, [\href{http://arxiv.org/abs/1601.06590}{{\tt arXiv:1601.06590}}].

\bibitem{Abdallah:2016ygi}
{\bf H.E.S.S.} Collaboration, H.~Abdallah et~al., {\it {Search for dark matter
  annihilations towards the inner Galactic halo from 10 years of observations
  with H.E.S.S}},  {\em Phys. Rev. Lett.} {\bf 117} (2016), no.~11 111301,
  [\href{http://arxiv.org/abs/1607.08142}{{\tt arXiv:1607.08142}}].

\bibitem{Meirose:2016pxn}
{\bf ATLAS} Collaboration, B.~Meirose, {\it {Overview of dark matter searches
  at the ATLAS experiment}},  {\em Int. J. Mod. Phys. Conf. Ser.} {\bf 43}
  (2016) 1660196.

\bibitem{Ahuja:2018bbj}
{\bf CMS} Collaboration, S.~Ahuja, {\it {Searches for Dark Matter with CMS}},
  {\em PoS} {\bf LHCP2018} (2018) 284.

\bibitem{Silveira:1985rk}
V.~Silveira and A.~Zee, {\it {SCALAR PHANTOMS}},  {\em Phys.\ Lett.\ B} {\bf
  161} (1985) 136--140.

\bibitem{Ma:1998dn}
E.~Ma, {\it {Pathways to naturally small neutrino masses}},  {\em Phys. Rev.
  Lett.} {\bf 81} (1998) 1171--1174,
  [\href{http://arxiv.org/abs/hep-ph/9805219}{{\tt hep-ph/9805219}}].

\bibitem{Krauss:2002px}
L.~M. Krauss, S.~Nasri, and M.~Trodden, {\it {A Model for neutrino masses and
  dark matter}},  {\em Phys. Rev. D} {\bf 67} (2003) 085002,
  [\href{http://arxiv.org/abs/hep-ph/0210389}{{\tt hep-ph/0210389}}].

\bibitem{Jueid:2020yfj}
A.~Jueid, S.~Nasri, and R.~Soualah, {\it {Searching for GeV-scale Majorana Dark
  Matter: inter spem et metum}},  {\em JHEP} {\bf 04} (2021) 012,
  [\href{http://arxiv.org/abs/2006.01348}{{\tt arXiv:2006.01348}}].

\bibitem{Liu:2021mhn}
J.~Liu, X.-P. Wang, and K.-P. Xie, {\it {Searching for lepton portal dark
  matter with colliders and gravitational waves}},
  \href{http://arxiv.org/abs/2104.06421}{{\tt arXiv:2104.06421}}.

\bibitem{Baum:2020gjj}
S.~Baum, P.~Sandick, and P.~Stengel, {\it {Hunting for scalar lepton partners
  at future electron colliders}},  {\em Phys. Rev. D} {\bf 102} (2020), no.~1
  015026, [\href{http://arxiv.org/abs/2004.02834}{{\tt arXiv:2004.02834}}].

\bibitem{Jueid:2020xx}
A.~Jueid and S.~Nasri, {\it {Dark Matter meets Unification}},  {\em {To be
  published}} (2021).

\bibitem{Branco:2011iw}
G.~C. Branco, P.~M. Ferreira, L.~Lavoura, M.~N. Rebelo, M.~Sher, and J.~P.
  Silva, {\it {Theory and phenomenology of two-Higgs-doublet models}},  {\em
  Phys. Rept.} {\bf 516} (2012) 1--102,
  [\href{http://arxiv.org/abs/1106.0034}{{\tt arXiv:1106.0034}}].

\bibitem{Kanemura:1993hm}
S.~Kanemura, T.~Kubota, and E.~Takasugi, {\it {Lee-Quigg-Thacker bounds for
  Higgs boson masses in a two doublet model}},  {\em Phys. Lett.} {\bf B313}
  (1993) 155--160, [\href{http://arxiv.org/abs/hep-ph/9303263}{{\tt
  hep-ph/9303263}}].

\bibitem{Akeroyd:2000wc}
A.~G. Akeroyd, A.~Arhrib, and E.-M. Naimi, {\it {Note on tree level unitarity
  in the general two Higgs doublet model}},  {\em Phys. Lett.} {\bf B490}
  (2000) 119--124, [\href{http://arxiv.org/abs/hep-ph/0006035}{{\tt
  hep-ph/0006035}}].

\bibitem{Arhrib:2015hoa}
A.~Arhrib, R.~Benbrik, J.~El~Falaki, and A.~Jueid, {\it {Radiative corrections
  to the Triple Higgs Coupling in the Inert Higgs Doublet Model}},  {\em JHEP}
  {\bf 12} (2015) 007, [\href{http://arxiv.org/abs/1507.03630}{{\tt
  arXiv:1507.03630}}].

\bibitem{Ginzburg:2010wa}
I.~F. Ginzburg, K.~A. Kanishev, M.~Krawczyk, and D.~Sokolowska, {\it {Evolution
  of Universe to the present inert phase}},  {\em Phys. Rev.} {\bf D82} (2010)
  123533, [\href{http://arxiv.org/abs/1009.4593}{{\tt arXiv:1009.4593}}].

\bibitem{Swiezewska:2012eh}
B.~Swiezewska and M.~Krawczyk, {\it {Diphoton rate in the inert doublet model
  with a 125 GeV Higgs boson}},  {\em Phys.\ Rev.\ D} {\bf 88} (2013), no.~3
  035019, [\href{http://arxiv.org/abs/1212.4100}{{\tt arXiv:1212.4100}}].

\bibitem{Jueid:2020rek}
A.~Jueid, J.~Kim, S.~Lee, S.~Y. Shim, and J.~Song, {\it {Phenomenology of the
  Inert Doublet Model with a global U(1) symmetry}},  {\em Phys. Rev. D} {\bf
  102} (2020), no.~7 075011, [\href{http://arxiv.org/abs/2006.10263}{{\tt
  arXiv:2006.10263}}].

\bibitem{Khachatryan:2016vau}
{\bf ATLAS, CMS} Collaboration, G.~Aad et~al., {\it {Measurements of the Higgs
  boson production and decay rates and constraints on its couplings from a
  combined ATLAS and CMS analysis of the LHC pp collision data at $ \sqrt{s}=7
  $ and 8 TeV}},  {\em JHEP} {\bf 08} (2016) 045,
  [\href{http://arxiv.org/abs/1606.02266}{{\tt arXiv:1606.02266}}].

\bibitem{Hahn:2000jm}
T.~Hahn, {\it {Automatic loop calculations with FeynArts, FormCalc, and
  LoopTools}},  {\em Nucl. Phys. Proc. Suppl.} {\bf 89} (2000) 231--236,
  [\href{http://arxiv.org/abs/hep-ph/0005029}{{\tt hep-ph/0005029}}].

\bibitem{Hahn:2000kx}
T.~Hahn, {\it {Generating Feynman diagrams and amplitudes with FeynArts 3}},
  {\em Comput. Phys. Commun.} {\bf 140} (2001) 418--431,
  [\href{http://arxiv.org/abs/hep-ph/0012260}{{\tt hep-ph/0012260}}].

\bibitem{Passarino:1978jh}
G.~Passarino and M.~Veltman, {\it {One Loop Corrections for e+ e- Annihilation
  Into mu+ mu- in the Weinberg Model}},  {\em Nucl.\ Phys.\ B} {\bf 160} (1979)
  151--207.

\bibitem{Sirunyan:2018owy}
{\bf CMS} Collaboration, A.~M. Sirunyan et~al., {\it {Search for invisible
  decays of a Higgs boson produced through vector boson fusion in proton-proton
  collisions at $\sqrt{s} =$ 13 TeV}},  {\em Phys. Lett. B} {\bf 793} (2019)
  520--551, [\href{http://arxiv.org/abs/1809.05937}{{\tt arXiv:1809.05937}}].

\bibitem{CMS-PAS-FTR-16-002}
{\bf CMS Collaboration} Collaboration, {\it {Projected performance of Higgs
  analyses at the HL-LHC for ECFA 2016}},  Tech. Rep. CMS-PAS-FTR-16-002, CERN,
  Geneva, 2017.

\bibitem{Cerri:2016bew}
O.~Cerri, M.~de~Gruttola, M.~Pierini, A.~Podo, and G.~Rolandi, {\it {Study the
  effect of beam energy spread and detector resolution on the search for Higgs
  boson decays to invisible particles at a future e$^+$ e$^-$ circular
  collider}},  {\em Eur. Phys. J. C} {\bf 77} (2017), no.~2 116,
  [\href{http://arxiv.org/abs/1605.00100}{{\tt arXiv:1605.00100}}].

\bibitem{Asner:2013psa}
D.~M. Asner et~al., {\it {ILC Higgs White Paper}},  in {\em {Proceedings, 2013
  Community Summer Study on the Future of U.S. Particle Physics: Snowmass on
  the Mississippi (CSS2013): Minneapolis, MN, USA, July 29-August 6, 2013}},
  2013.
\newblock \href{http://arxiv.org/abs/1310.0763}{{\tt arXiv:1310.0763}}.

\bibitem{CEPCStudyGroup:2018ghi}
{\bf CEPC Study Group} Collaboration, M.~Dong et~al., {\it {CEPC Conceptual
  Design Report: Volume 2 - Physics \textbackslash{}\& Detector}},
  \href{http://arxiv.org/abs/1811.10545}{{\tt arXiv:1811.10545}}.

\bibitem{Selvaggi:2018obt}
M.~Selvaggi, {\it {Higgs measurements at the FCC-hh}},  {\em PoS} {\bf
  ICHEP2018} (2019) 684.

\bibitem{Adam:2013mnn}
{\bf MEG} Collaboration, J.~Adam et~al., {\it {New constraint on the existence
  of the $\mu^+ \to e^+\gamma$ decay}},  {\em Phys. Rev. Lett.} {\bf 110}
  (2013) 201801, [\href{http://arxiv.org/abs/1303.0754}{{\tt
  arXiv:1303.0754}}].

\bibitem{Aubert:2009ag}
{\bf BaBar} Collaboration, B.~Aubert et~al., {\it {Searches for Lepton Flavor
  Violation in the Decays tau+- ---> e+- gamma and tau+- ---> mu+- gamma}},
  {\em Phys. Rev. Lett.} {\bf 104} (2010) 021802,
  [\href{http://arxiv.org/abs/0908.2381}{{\tt arXiv:0908.2381}}].

\bibitem{Ahriche:2017iar}
A.~Ahriche, A.~Jueid, and S.~Nasri, {\it {Radiative neutrino mass and Majorana
  dark matter within an inert Higgs doublet model}},  {\em Phys. Rev.} {\bf
  D97} (2018), no.~9 095012, [\href{http://arxiv.org/abs/1710.03824}{{\tt
  arXiv:1710.03824}}].

\bibitem{Ahriche:2018ger}
A.~Ahriche, A.~Arhrib, A.~Jueid, S.~Nasri, and A.~de~La~Puente, {\it
  {Mono-Higgs Signature in the Scotogenic Model with Majorana Dark Matter}},
  {\em Phys. Rev. D} {\bf 101} (2020), no.~3 035038,
  [\href{http://arxiv.org/abs/1811.00490}{{\tt arXiv:1811.00490}}].

\bibitem{Conte:2012fm}
E.~Conte, B.~Fuks, and G.~Serret, {\it {MadAnalysis 5, A User-Friendly
  Framework for Collider Phenomenology}},  {\em Comput. Phys. Commun.} {\bf
  184} (2013) 222--256, [\href{http://arxiv.org/abs/1206.1599}{{\tt
  arXiv:1206.1599}}].

\bibitem{Conte:2014zja}
E.~Conte, B.~Dumont, B.~Fuks, and C.~Wymant, {\it {Designing and recasting LHC
  analyses with MadAnalysis 5}},  {\em Eur. Phys. J. C} {\bf 74} (2014), no.~10
  3103, [\href{http://arxiv.org/abs/1405.3982}{{\tt arXiv:1405.3982}}].

\bibitem{Conte:2018vmg}
E.~Conte and B.~Fuks, {\it {Confronting new physics theories to LHC data with
  MADANALYSIS 5}},  {\em Int. J. Mod. Phys.} {\bf A33} (2018), no.~28 1830027,
  [\href{http://arxiv.org/abs/1808.00480}{{\tt arXiv:1808.00480}}].

\bibitem{Fuks:2021wpe}
B.~Fuks et~al., {\it {Proceedings of the second MadAnalysis 5 workshop on LHC
  recasting in Korea}},  \href{http://arxiv.org/abs/2101.02245}{{\tt
  arXiv:2101.02245}}.

\bibitem{Araz:2020dlf}
J.~Y. Araz and B.~Fuks, {\it {Implementation of the ATLAS-SUSY-2018-32 analysis
  (sleptons and electroweakinos with two leptons and missing transverse energy;
  139~fb\ensuremath{-}1)}},  {\em Mod. Phys. Lett. A} {\bf 36} (2020), no.~01
  2141005.

\bibitem{Ambrogi:2018jqj}
F.~Ambrogi, C.~Arina, M.~Backovic, J.~Heisig, F.~Maltoni, L.~Mantani,
  O.~Mattelaer, and G.~Mohlabeng, {\it {MadDM v.3.0: a Comprehensive Tool for
  Dark Matter Studies}},  {\em Phys. Dark Univ.} {\bf 24} (2019) 100249,
  [\href{http://arxiv.org/abs/1804.00044}{{\tt arXiv:1804.00044}}].

\bibitem{Baer:2013cma}
H.~Baer, T.~Barklow, K.~Fujii, Y.~Gao, A.~Hoang, S.~Kanemura, J.~List, H.~E.
  Logan, A.~Nomerotski, M.~Perelstein, et~al., {\it {The International Linear
  Collider Technical Design Report - Volume 2: Physics}},
  \href{http://arxiv.org/abs/1306.6352}{{\tt arXiv:1306.6352}}.

\bibitem{Alwall:2014hca}
J.~Alwall, R.~Frederix, S.~Frixione, V.~Hirschi, F.~Maltoni, O.~Mattelaer,
  H.~S. Shao, T.~Stelzer, P.~Torrielli, and M.~Zaro, {\it {The automated
  computation of tree-level and next-to-leading order differential cross
  sections, and their matching to parton shower simulations}},  {\em JHEP} {\bf
  07} (2014) 079, [\href{http://arxiv.org/abs/1405.0301}{{\tt
  arXiv:1405.0301}}].

\bibitem{Sjostrand:2014zea}
T.~Sj{\"o}strand, S.~Ask, J.~R. Christiansen, R.~Corke, N.~Desai, P.~Ilten,
  S.~Mrenna, S.~Prestel, C.~O. Rasmussen, and P.~Z. Skands, {\it {An
  Introduction to PYTHIA 8.2}},  {\em Comput. Phys. Commun.} {\bf 191} (2015)
  159--177, [\href{http://arxiv.org/abs/1410.3012}{{\tt arXiv:1410.3012}}].

\bibitem{deFavereau:2013fsa}
{\bf DELPHES 3} Collaboration, J.~de~Favereau, C.~Delaere, P.~Demin,
  A.~Giammanco, V.~Lemaître, A.~Mertens, and M.~Selvaggi, {\it {DELPHES 3, A
  modular framework for fast simulation of a generic collider experiment}},
  {\em JHEP} {\bf 02} (2014) 057, [\href{http://arxiv.org/abs/1307.6346}{{\tt
  arXiv:1307.6346}}].

\bibitem{Cacciari:2008gp}
M.~Cacciari, G.~P. Salam, and G.~Soyez, {\it {The Anti-k(t) jet clustering
  algorithm}},  {\em JHEP} {\bf 04} (2008) 063,
  [\href{http://arxiv.org/abs/0802.1189}{{\tt arXiv:0802.1189}}].

\bibitem{Cacciari:2011ma}
M.~Cacciari, G.~P. Salam, and G.~Soyez, {\it {FastJet User Manual}},  {\em Eur.
  Phys. J.} {\bf C72} (2012) 1896, [\href{http://arxiv.org/abs/1111.6097}{{\tt
  arXiv:1111.6097}}].

\bibitem{Aoki:2010tf}
M.~Aoki and S.~Kanemura, {\it {Probing the Majorana nature of TeV-scale
  radiative seesaw models at collider experiments}},  {\em Phys.\ Lett.\ B}
  {\bf 689} (2010) 28--35, [\href{http://arxiv.org/abs/1001.0092}{{\tt
  arXiv:1001.0092}}].

\bibitem{Read:2002hq}
A.~L. Read, {\it {Presentation of search results: The CL(s) technique}},  {\em
  J. Phys.} {\bf G28} (2002) 2693--2704. ,11 (2002).

\end{thebibliography}\endgroup


\end{document}